\documentclass[conference]{IEEEtran} 
\IEEEoverridecommandlockouts
\usepackage{cite}
\usepackage{amsmath,amssymb,amsfonts}
\usepackage{graphicx}
\usepackage{textcomp}
\usepackage{booktabs}
\usepackage{hyperref}

\usepackage{algorithm}
\usepackage{algpseudocode}
\usepackage{tabularx}
\usepackage{xcolor}
\usepackage{makecell}

\def\BibTeX{{\rm B\kern-.05em{\sc i\kern-.025em b}\kern-.08em
    T\kern-.1667em\lower.7ex\hbox{E}\kern-.125emX}}
\begin{document}

\title{A Modular and Scalable Simulator for Connected-UAVs Communication in 5G Networks \\
}

\author{
Yong Su\textsuperscript{1}, 
Yiyi Chen\textsuperscript{2}, 
Shenghong Yi\textsuperscript{1},
Hui Feng\textsuperscript{1}\textsuperscript{*} \thanks{*Corresponding author: Hui Feng}, 
Yuedong Xu\textsuperscript{3}, 
Wang Xiang\textsuperscript{4}, 
Bo Hu\textsuperscript{1} \\
\textsuperscript{1}College of Future Information Technology, Fudan University, Shanghai 200433, China \\
\textsuperscript{2}College of Intelligent Robotics and Advanced Manufacturing, Fudan University, Shanghai 200433, China\\ 
\textsuperscript{3}College of Computer Science and Artificial Intelligence, Fudan University, Shanghai 200438, China \\
\textsuperscript{4}Informatization Office, Fudan University, Shanghai 200433, China\\
\{suy21, yiyichen21, shyi22\}@m.fudan.edu.cn, \{hfeng, ydxu, xiangw, bohu\}@fudan.edu.cn\\
}

\maketitle

\begin{abstract}
Cellular-connected UAV systems have enabled a wide range of low-altitude aerial services. 
However, these systems still face many challenges, such as frequent handovers and the inefficiency of traditional transport protocols. 
To better study these issues, we develop a modular and scalable simulation platform specifically designed for UAVs communication leveraging the research ecology in wireless communication of MATLAB. 
The platform supports flexible 5G NR node deployment, customizable UAVs mobility models, and multi-network-interface extensions. 
It also supports multiple transport protocols including TCP, UDP, QUIC, etc., allowing to investigate how different transport protocols affect UAVs communication performance.
In addition, the platform includes a handover management module, enabling the evaluation of both traditional and learning-based handover strategies. 
Our platform can serve as a testbed for the development and evaluation of advanced transmission strategies in cellular-connected UAV systems.

\end{abstract}


\section{Introduction}

\begin{figure}[htbp]
\centerline{\includegraphics[width=1\linewidth]{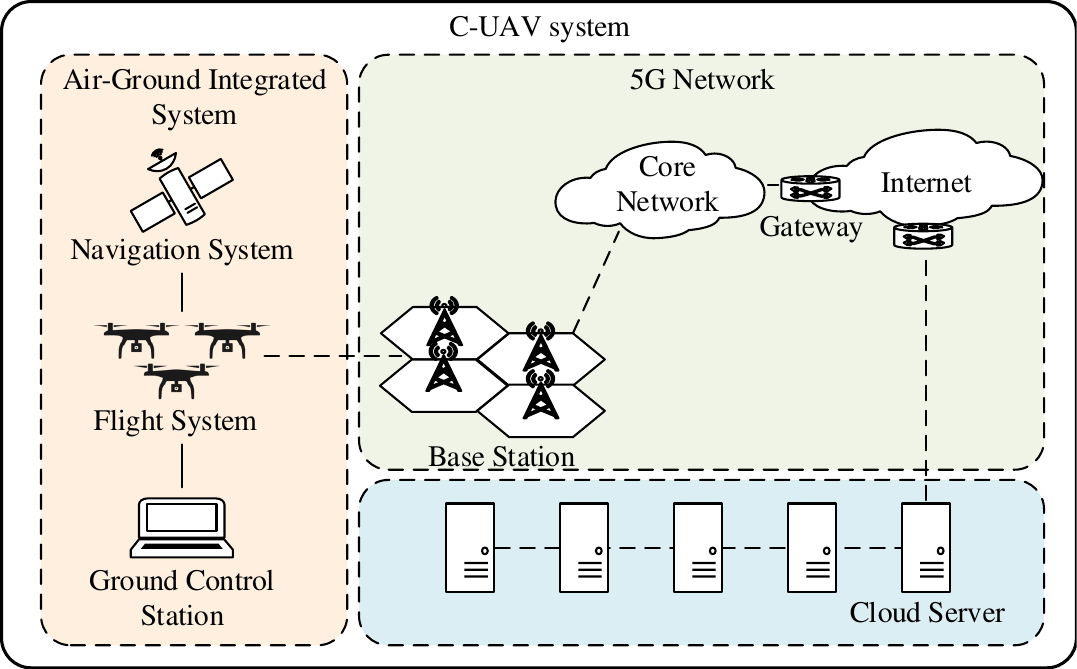}}
\caption{System architecture of C-UAVs.}
\label{fig:connected-UAV-systems}
\end{figure}

Cellular-connected UAVs (C-UAVs) are UAVs integrated into cellular networks as aerial user equipments (UEs), supporting applications such as logistics, agriculture, and emergency response\cite{Qazzaz2024NonTerrestrial}.
A typical system involves UAVs for data collection and ground control stations (or cloud servers) for command and data exchange\cite{hentati2022simulation}.
The architecture among these components of a C-UAV system is shown in the Fig. \ref{fig:connected-UAV-systems}.
Compared with 4G, 5G offers higher uplink capacity, lower latency, and support for massive machine-type communications, making it suitable for real-time control, video streaming, and reliable connectivity\cite{Babafaa2024Survey}.
However, C-UAV communications still face challenges: downtilted antennas and side-lobe coverage cause weak and unstable links\cite{Benzaghta2025Cellular}, line-of-sight interference degrades reliability at higher altitudes\cite{ullah2020cognition,Zhou2024Handover}, and high mobility leads to frequent handovers, including ping-pong effects\cite{Irshad2024Mobility}. Furthermore, traditional transport protocols like TCP are highly sensitive to packet loss and latency variations\cite{xu2021deep, ayass2022uavchallenge}, resulting in degraded QoS and QoE.

\begin{table*}[!htb]
\centering
\caption{Comparison of UAVs Communication Simulation Tools}
\label{tab:network-simulators-comparious}

\begin{tabular}{@{}p{0.9in}p{2in}p{2in}>{\raggedright\arraybackslash}p{2in}@{}} 
\toprule
\textbf{Tools} & \textbf{UAV Support} & \textbf{Strengths} & \textbf{Limitations} \\ 
\midrule

ns-3 & 
\makecell[l]{Supports UAV;\\ Network topology modeling;} & 
\makecell[l]{Modular architecture;\\ Highly accurate simulations;\\ Active developer community;} & 
\makecell[l]{Limited support for UAV characteristics;\\ Steep learning curve;} \\
\\
OMNeT++ & 
\makecell[l]{UAV simulation via UAVSim\\or custom module;} & 
\makecell[l]{Rich GUI;\\ High-precision event-driven simulation;} & 
\makecell[l]{Complex configuration;\\ Limited native UAV support;\\ Requires extensions for mobility models;} \\
\\
EXata & 
\makecell[l]{Supports virtual node-based\\ UAV modeling;} & 
\makecell[l]{High simulation speed;\\ Real-time interfacing with external systems;} & 
\makecell[l]{Not open-source;\\ Expensive licensing;\\ Limited flexibility for customization;} \\

\bottomrule
\end{tabular}
\end{table*}

Due to these challenges, C-UAVs require systematic investigation.
Conducting field experiments with large-scale UAVs is expensive and difficult to control. 
Simulation becomes an efficient approach for studying and validating communication strategies in C-UAV systems. 
Although several simulators exist for wireless research, few address UAV-specific needs.
Table~\ref{tab:network-simulators-comparious} summarizes key features and limitations of widely used simulators in UAVs-related studies\cite{hentati2022simulation}. 
Gaps remain: 
limited support for emerging protocols and AI algorithms\cite{jiang2024survey}; insufficient UAV-specific modeling; lack of open-source access.
Furthermore, most simulators fail to model multi-interface configurations essential for reliable connectivity. 

To address these challenges, there is a need for a modular, scalable, and open-source simulation platform capable of comprehensive modeling of C-UAV system. 
Such a platform should: 
(1) represent UAV-specific characteristics; 
(2) support scalable network topology configurations, particularly multi-interface setups for multi-operator connectivity studies; 
(3) support full protocol stack and end-to-end communication; 
(4)  include a handover management module that supports event-based algorithms and integrates learning-based strategies for flexible handover policy evaluation. 
Crucially, a modular architecture for these core functions can ensure maintainability, customization, and adaptability to evolving C-UAVs scenarios.

In this paper, we design and develop a simulation platform specifically tailored for C-UAV systems, built upon MATLAB\footnote{The code of this simulator is available at: \url{https://github.com/suyong-123/5G_C-UAV_Matlab_Simulator}.}.
MATLAB is the widely used platform for wireless communication algorithm development, particularly for the physical and link layers. Its rich ecosystem provides extensive resources supporting research on C-UAVs. Moreover, MATLAB offers extensive interfaces, excellent protocol extensibility, and a highly user-friendly programming environment. 
However, despite its comprehensive support for 5G standards, dedicated simulation capabilities for the C-UAVs remain lacking. Therefore, leveraging the foundation provided by MATLAB 5G Toolbox, we develop this platform to specifically address the simulation needs of C-UAV systems.

The proposed simulation platform employs an end-to-end system-level modeling, integrating UAVs mobility models, air-to-ground propagation characteristics, full protocol stack implementation, handover procedures, and multi-interface terminal support. 
Unlike existing simulators such as ns-3 or OMNeT++, which are primarily designed for terrestrial networks and require complex extensions for UAVs or emerging transport protocols, our platform offers a unified, modular environment specifically tailored for aerial networks, enabling realistic evaluation of advanced mobility and transmission strategies.
Our main contributions are summarized as follows:
\begin{itemize}
\item \textbf{Scalable Simulation Setting}: The platform supports flexible configuration of 5G NR nodes such as position, power, and frequency settings, and allows optional use of multiple network interfaces. UAVs mobility is modeled using either randomized movement patterns within a bounded 3D space or predefined trajectories. These features collectively allow constructing diverse simulation environments. 

    \item \textbf{Support for Multiple Transport Protocols}: The platform supports multiple transport protocols, including TCP, UDP, and QUIC, enabling evaluation of their performance in C-UAV systems.

    \item \textbf{Extensible Handover Management}: The platform implements an extensible handover management module supporting both event-based algorithms (e.g., A3 events) and reinforcement learning strategies, including our proposed Cross-layer DQN. 
\end{itemize}

The platform encapsulates these key functionalities into modular, reusable classes. This architecture supports evaluation of transmission strategies for C-UAVs in scalable scenarios, with a focus on connection quality, throughput, and handover frequency in low-altitude environments. 

The rest of the paper is organized as follows.
Section II presents the design of the proposed simulation platform.
Section III describes the experimental setup and evaluates the platform’s functionality.
Section IV concludes the paper and discusses future research directions.

\section{SYSTEM DESIGN}

\begin{figure}[!ht]
\centerline{\includegraphics[width=1\linewidth]{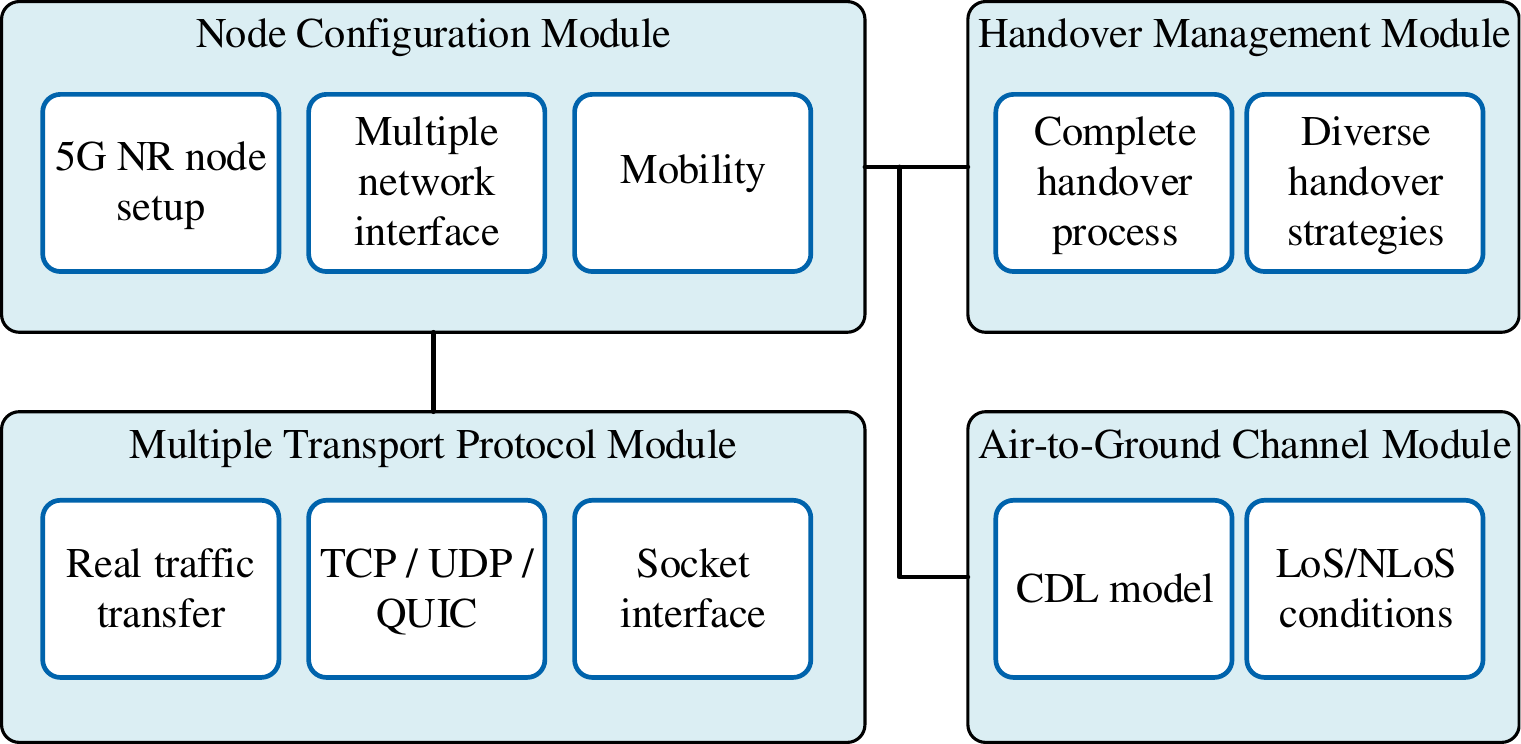}}
\caption{The structure diagram of proposed 5G C-UAVs simulator. }
\label{fig:Our-simulator}
\end{figure}

The proposed simulation platform’s complete framework is illustrated in Fig. \ref{fig:Our-simulator}, which we divide into four key modules: node configuration, channel model, multiple transport protocols integration, and handover management.

\subsection{Node Configuration}

The simulation platform provides a comprehensive modeling for key network elements, including base stations (gNBs), UEs simulating UAVs, and their additional functionalities such as mobility behaviors, multi-network interfaces, and built-in 5G access-layer protocol stacks. 
The platform implements a flexible deployment mechanism that allows researchers to customize node parameters and mobility settings.

All network entities, including gNBs and UEs, are modeled as configurable nodes. Each node can be assigned parameters such as spatial position, carrier frequency, channel bandwidth, antenna configuration, receive gain, and duplexing mode. The platform supports both manual specification and randomized placement of nodes, enabling flexible experimentation with scalable network topologies and coverage scenarios.

In addition to static positioning, each UE can be configured with either a 3D random waypoint mobility model or customized trajectory patterns.
The 3D random waypoint mobility model (provided by MATLAB) enables UAVs to move within predefined spatial boundaries with configurable velocity and hovering parameters.
The proposed simulation platform extends this capability by adding deterministic trajectory support, allowing users to design fixed routes through a dedicated path planner.  

To support multi-connectivity, the simulation platform enables heterogeneous network deployment with two sets of gNBs operating at 2.6 GHz and 3.5 GHz to emulate distinct operator domains. 
The design instantiates two or more UE entities with synchronized mobility models, emulating a single terminal featuring multi-network interfaces. Each UE connects to distinct gNB groups operating on different frequency bands, enabling a single mobile terminal to simultaneously access multiple operators' networks.

Each node in the simulation platform incorporates  a full-stack 5G NR access protocol implementation provided by MATLAB 5G Toolbox, including the physical layer (PHY), medium access control (MAC), and radio link control (RLC) layers. This layered architecture enables accurate modeling of scheduling, HARQ, retransmission, and buffer management mechanisms, which is essential for evaluating cross-layer transmission strategies.

\subsection{ Channel Model}

In terms of channel modeling, this work adopts the standardized Clustered Delay Line (CDL) model provided by MATLAB 5G Toolbox. However, this model is primarily designed for ground users and does not fully capture the unique propagation characteristics experienced by aerial users at higher altitudes, such as increased line-of-sight probability, reduced multipath richness, and altitude-dependent path loss.  

To better reflect UAV-specific behaviors, this study incorporates aerial-user-specific modeling approaches defined in 3GPP TR 36.777 \cite{3gpptr36777}. In particular, we consider the Urban Micro Aerial Vehicle scenario, which accounts for low-altitude UAV deployments with environment- and altitude-dependent channel parameters, thus accurately representing the propagation conditions that UAVs encounter in realistic operations.

\subsection{Multiple Transport Protocols Integration }

\begin{figure}[htbp]
\centerline{\includegraphics[width=1\linewidth]{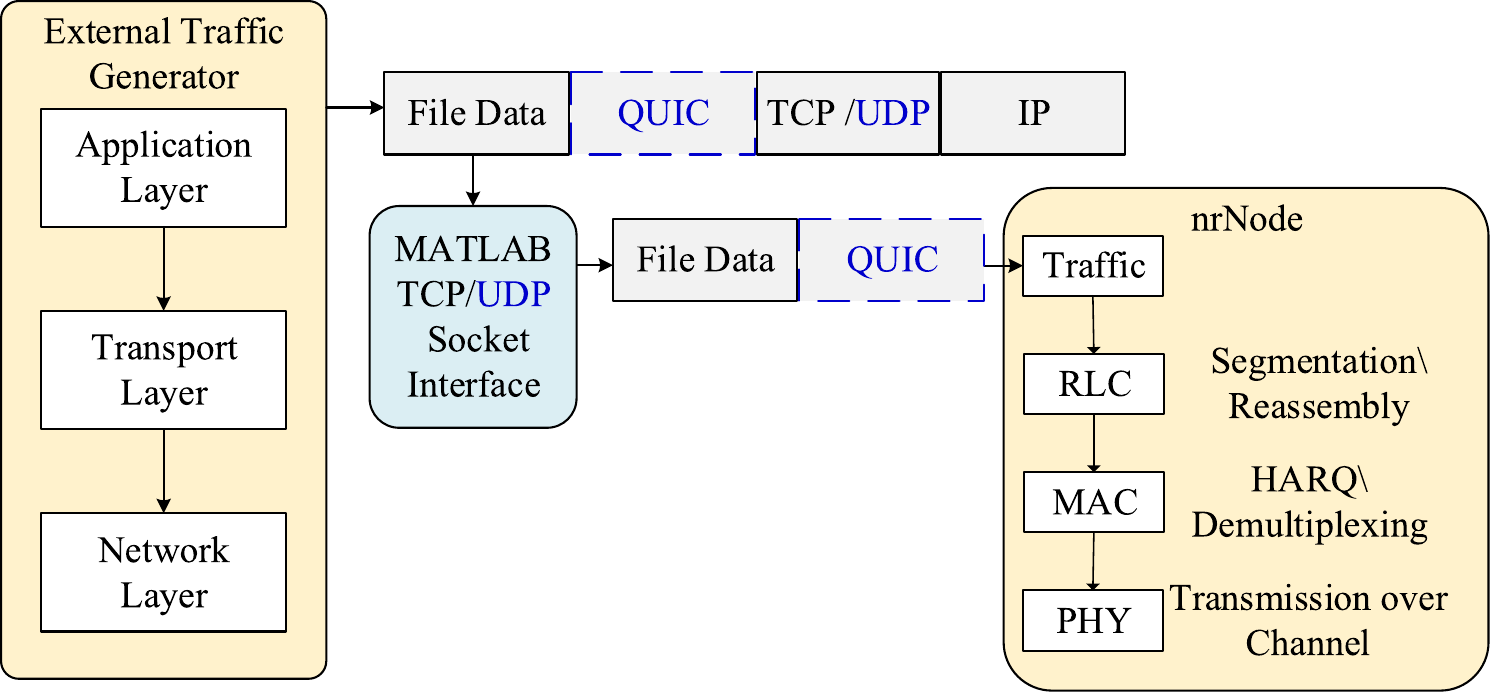}}
\caption{Protocol stack integration of external traffic generator with MATLAB 5G simulation. }
\label{fig:stack}
\end{figure}

Transport protocols play a vital role in end-to-end communication performance. 
However existing researches rarely explore the influence of different transport protocols on the performance of C-UAV systems. 
To address this gap, our simulation platform supports multiple transport protocols including TCP, UDP, and QUIC, enabling systematic and comparative studies across the entire network protocol stack.

While the MATLAB 5G Toolbox provides detailed 5G NR protocol stack, its built-in traffic layer only supports synthetic traffic generation and lacks support for actual upper-layer protocols such as transport and application layers. 
To emulate realistic end-to-end communication behavior, we use external traffic sources that generate real application-layer data encapsulated with standard transport protocols.

As shown in Fig. \ref{fig:stack}, we implement a socket-based interface between external applications and MATLAB. 
This diagram depicts the workflow for transferring data packets from external applications into the MATLAB simulation network from a protocol stack perspective. It illustrates how real application layer data generated by external programs is encapsulated from the application layer to the network layer, then transmitted via sockets.  
On the MATLAB side, the  \verb|tcpclient| or  \verb|udpport| object receives these packets via event-driven callbacks. 
Upon reception, the data packet is directly assigned to the  \verb|Traffic| layer of the corresponding network node as the payload to be transmitted. 
This figure also details the operational mechanisms of each layer within the MATLAB nrNode's built-in stack, showing how data is processed hierarchically. 
All socket communication leverages the underlying operating system's network drivers. 

\begin{figure}[htbp]
\centerline{\includegraphics[width=1\linewidth]{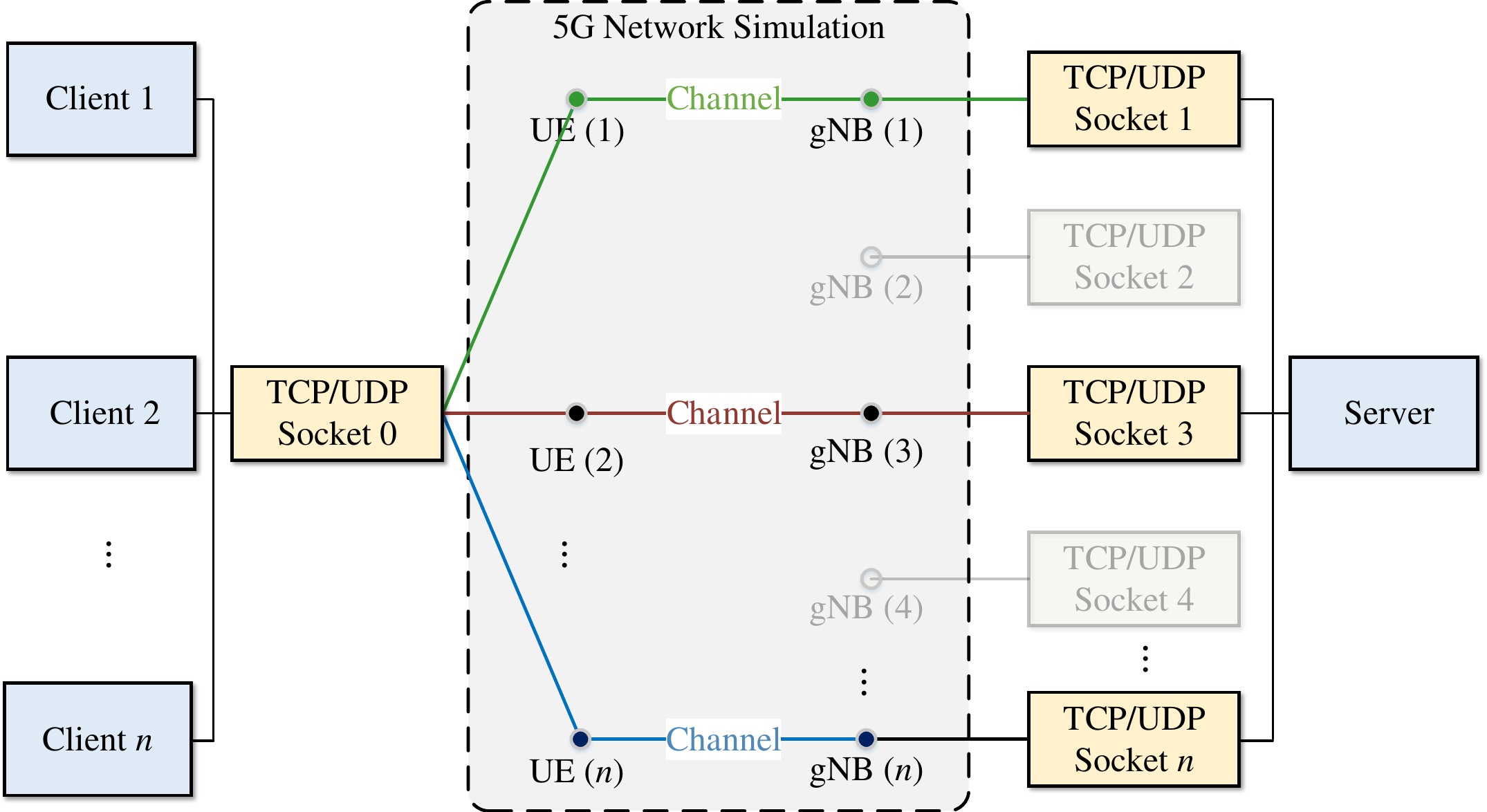}}
\caption{External traffic transfer via MATLAB socket interface. }
\label{fig:multi}
\end{figure}

Building upon this foundation, Fig.  \ref{fig:multi} showcases the scalability of our platform through a socket-based interface that enables concurrent bidirectional data exchange among multiple external clients and MATLAB. 
Each incoming packet is uniquely identified using custom headers, allowing the system to route traffic to dedicated UE instances. As illustrated, this architecture supports parallel processing of diverse traffic profiles across multiple UEs, with each flow traversing the complete 5G protocol stack to respective gNBs. Upon receiving a packet, the gNB instantaneously relays it to the server-side via its bound socket.
Downlink traffic follows an inverse path, with packets dynamically routed to targeted UE based on header metadata. This design demonstrates the platform's ability to scale horizontally, handling increasing client loads while maintaining protocol fidelity and performance metrics. 

Traditional transport protocols such as TCP usually struggle to meet the communication requirements of high mobility UAVs scenarios. 
The proposed simulation platform adds support for QUIC, a transport protocol proposed by Google and standardized by the IETF as the foundation of HTTP/3\cite{quic}. 
QUIC offers several advantages that make it particularly suitable for UAVs communication, including low connection establishment latency, stream multiplexing without head-of-line blocking, support for connection migration, and seamless path switching\cite{Lei2024quic}.

However, MATLAB does not support QUIC, and successful communication between a QUIC client and server requires both endpoints to implement the QUIC protocol to complete the handshake. 
To bridge this gap, we design a bridging architecture that leverages QUIC’s underlying UDP transport to interface MATLAB with external QUIC-Go implementations. As shown in the blue-shaded area of Fig. \ref{fig:stack}, MATLAB's UDP interface receives QUIC packets encapsulated within UDP datagrams.

\begin{figure}[htbp]
\centerline{\includegraphics[width=1\linewidth]{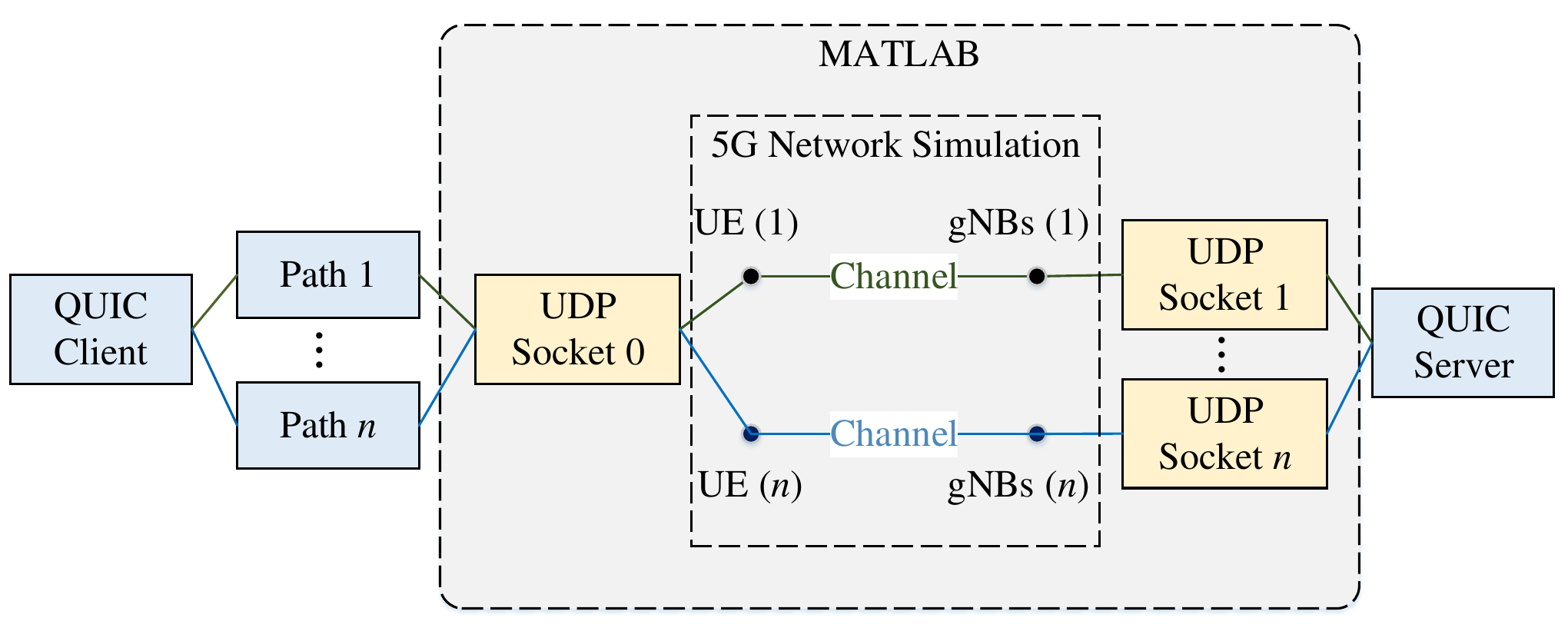}}
\caption{Multi-Path QUIC with MATLAB 5G network simulation. }
\label{fig:mpquic-matlab}
\end{figure}

To support multi-path transmission based on MP-QUIC, the framework is extended to handle multiple QUIC paths. 
Specifically, as illustrated in Fig. \ref{fig:mpquic-matlab}, 
all uplink packets from the external MP-QUIC client are received by a single listening socket and dispatched to appropriate UE based on path-identifying information (source IP/port). 
After traversing the full 5G protocol stack, the packet is delivered to the gNBs group serving the UE on corresponding frequency bands .
This socket configuration effectively enables asymmetric communication: a many-to-one pattern in the uplink, where multiple client paths send data to a common server port, and a one-to-many pattern in the downlink, where the server responds to each path via distinct destination ports.

\subsection{Handover Management}
The proposed simulation platform integrates a modular and extensible handover management framework. This framework supports both traditional rule-based policies and learning-based decision-making, and is fully encapsulated within a dedicated class called \verb|handoverManager|, which is assigned to each UE.

Handover is a fundamental mobility management procedure in cellular networks\cite{shayea2022handover}, designed to ensure the continuity and quality of communication services as UE moves across the coverage areas of different gNBs. 
A typical handover process consists of three stages: measurement, decision, and execution\cite{haghrah2023survey}.
(i) In the measurement stage, the UE periodically measures the signal quality of neighboring cells based on network configuration and reports relevant indicators, such as the Reference Signal Received Power (RSRP) or the Signal-to-Interference-plus-Noise Ratio (SINR).
(ii) In the decision stage, the source gNB determines whether a handover should be triggered based on the measurement report and pre-defined event conditions, and selects the optimal target cell. 
(iii) In the execution stage, the UE disconnects from the serving gNB and completes the access procedure with the target gNB, thus completing the wireless link migration.

We implement a modular \verb|handoverManager| class that encapsulates the full decision-making and execution process.
The \verb|handoverManager| class performs the following:

\begin{itemize}
    \item \textbf{Collect SINR measurements from packet reception events.} Sounding Reference Signal (SRS) transmissions provide uplink channel quality information from each gNB such as SINR. 
However, MATLAB does not support SINR measurement from all surrounding gNBs—each UE can only obtain the SINR value from its currently connected serving gNB. 
To enable SINR measurement from neighboring (non-serving) gNBs, we establish dedicated uplink links between the UE and each non-serving gNB specifically for SRS transmission. This mechanism is encapsulated in the function.
Specifically, the UE transmits its uplink configuration parameters to all nearby gNBs. Each gNB then configures the necessary SRS reception structures within its MAC and PHY layers, enabling it to listen for and decode incoming SRS transmissions. In this way, non-serving gNBs can periodically receive SRS signals from the UE and compute the corresponding SINR based on the received signal power. This provides essential channel quality information to support subsequent handover decisions. 

    \item \textbf{Maintain a sliding average of recent measurements.} Instantaneous SINR values may fluctuate due to fast fading and transient interference. We use a short-term average to smooth out noise and ensure more stable and reliable handover decisions. 

    \item \textbf{Handover Strategies.} \verb|handoverManager| supports multiple configurable handover strategies, including both rule-based and learning-based policies. 
Among the supported strategies, the A3 event-based handover, widely adopted in 3GPP specifications serves as the baseline mechanism in our simulation.
 
    \item \textbf{Execute handover. }when condition is met, UE will disconnect from the source gNB, reconnect to the target gNB, and reset associated traffic flows. 
\end{itemize}

\begin{figure}[htbp]
\centerline{\includegraphics[width=1\linewidth]{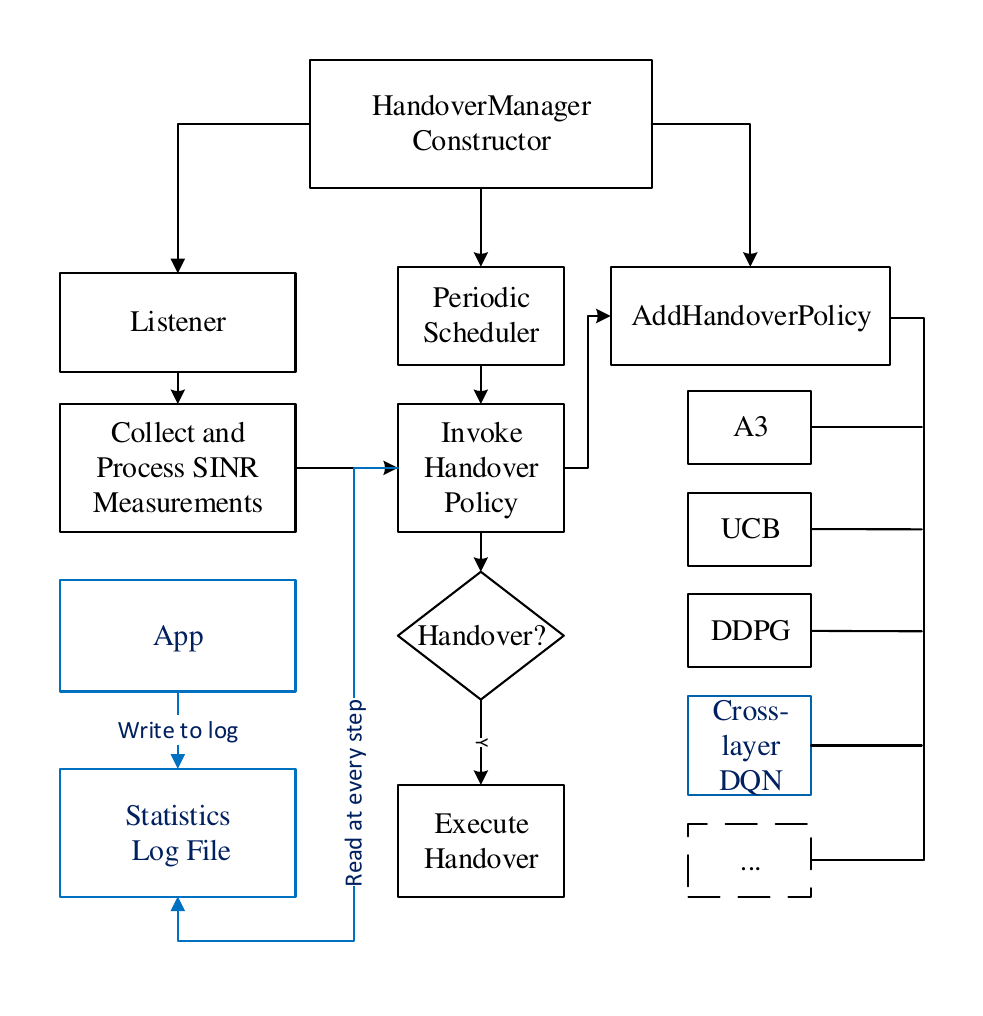}}
\caption{ Workflow of handover management. }
\label{fig:handovermanager_class_flow}
\end{figure}

The internal logic and execution flow of the \verb|handoverManager| class is illustrated in Fig. \ref{fig:handovermanager_class_flow}, which highlights the interaction between signal measurement and handover execution routines.

\section{DRL-ENABLED UAV HANDOVER ALGORITHM  }

In recent years, reinforcement learning (RL) has gained increasing attention for optimizing mobility management\cite{ZAID2024review}.
Notable examples include the SMART framework by Sun et al. \cite{sun2018smart}, which applied UCB and integer programming to reduce handover frequency in mmWave networks. Similarly, Kwong et al. \cite{kwong2022ddpg} proposed a DDPG-based scheme for dynamically adjusting handover margins in ultra-dense 5G deployments.
However, these studies are inherently constrained by their reliance on traditional TCP/IP transport protocols, failing to leverage the unique features of QUIC. 
Few of the existing RL-based handover algorithms are tailored to exploit QUIC's protocol characteristics in upper layers.

To bridge this gap, the simulation platform introduces a cross-layer deep reinforcement learning (DRL) framework, which addresses the critical limitation of prior work where handover decisions were decoupled from transport-layer behavior by fostering synergistic interaction between mobility management and transport protocol features. The platform enables cross-layer DRL in C-UAVs scenarios by providing physical layer signal quality metrics and transport layer performance feedback (e.g., throughput). 

We instrument the \verb|ReceivedAck| function to log key metrics—such as RTT, packet losses, and total transmitted bytes—whenever RTT is updated.
These statistics are appended to a log file during runtime. The \verb|handoverManager| class in MATLAB periodically parses this file to retrieve the latest throughput information, which is then normalized and incorporated into the DRL state space.
This mechanism allows the DRL agent to make mobility decisions not only based on physical layer SINR but also on application layer throughput performance.

We now detail the modeling structure of our cross-layer DRL framework, including the state space, action space, reward function and optimized algorithm.

\textbf{State space:}

The state is designed to capture both radio-link quality and transport layer performance, enabling cross-layer decision-making. Specifically, the agent observes:
\[
\mathbf{s} = \left\{ \text{SINR}_{1}, \text{SINR}_{2}, \dots, \text{SINR}_{|B|}, \mathbf{1}_{\text{current}}, \Delta_{\text{best}}, T, \tau, H \right\}
\]
where  $\text{SINR}_{j}$ denotes the normalized uplink SINR from the \textit{j}-th gNB, $|B|$ is the total number of candidate base stations, $\textbf{1}_{current}$ is the one-hot encoding indicating the currently connected gNB, and $T$ is the normalized uplink throughput parsed from the external QUIC log file. 
 $\Delta_{\text{best}}$ represents the normalized SINR gap between the currently connected gNB and the best available gNB. $\tau$ is the normalized time since the last handover, and $H$ is the normalized number of handovers executed so far. 
 This structure ensures that the agent can learn from both link-layer and end-to-end performance indicators.

\textbf{Action space:}

The action space corresponds to the set of all available gNBs in the UE’s coverage area. At each decision step, the agent selects one base station as the potential serving node:
\[
a = [b_j], \quad b_j \in B
\]
 where $b_{j}$ is the index of the next serving base station. If the chosen $b_j$ is already the currently connected gNB, no handover is performed. 
This discrete action space directly maps to the set of feasible handover targets for each UE.

\textbf{Reward function:}

The reward is designed to jointly maximize throughput and signal quality, while penalizing unnecessary handovers:
\[
r = w_1 \cdot \text{SINR}_{\text{target}} + w_2 \cdot T - w_3 \cdot \text{HO}_{\text{cost}}
\]
where $\text{SINR}_\text{target}$ is the normalized SINR of the target gNB,  $\text{HO}_\text{cost}$ = 1, if a handover occurs, and 0 otherwise. 
The weights $w_1$, $w_2$, and $w_3$ are tunable parameters controlling the trade-off between performance and stability.
By incorporating throughput into the reward, the agent learns to avoid handovers that yield only SINR improvements at the cost of end-to-end performance degradation. This design helps the agent trade off between signal quality and transmission stability.

\textbf{DRL Algorithm:}

Traditional tabular RL algorithms are impractical for our high-dimensional, cross-layer handover scenario\cite{thrun2000reinforcement}. To address this limitation, we adopt a DRL approach, which leverages neural networks to approximate the action-value function and enables generalization to unseen states. Among DRL methods, DQN \cite{Hafiz2022DQNsurvey} is a widely used off-policy algorithm that incorporates experience replay and target networks. However, DQN is known to suffer from overestimation bias, which can lead to unstable training and suboptimal policies. To mitigate this issue, we employ Double DQN, which decouples action selection and evaluation by maintaining two separate networks—an online network for choosing actions and a target network for estimating their values—thus providing more reliable learning targets. 

Building upon the standard Double DQN, we introduce several enhancements to further improve learning stability and decision performance. Specifically, we incorporate the dueling network architecture, which separately estimates the state value and the advantage of each action, allowing the agent to better distinguish valuable actions in varying contexts. To address sparse reward signals and accelerate learning, we adopt N-step temporal difference learning, enabling the agent to consider multiple future rewards in each update. This enhanced Double DQN agent is trained offline through episodic interactions with the simulated environment and is integrated into the online handover decision module for real-time evaluation.

\section{SYSTEM EVALUATION}

\subsection{C-UAV System Mobility}

\begin{figure}[htbp]
\centerline{\includegraphics[width=1\linewidth]{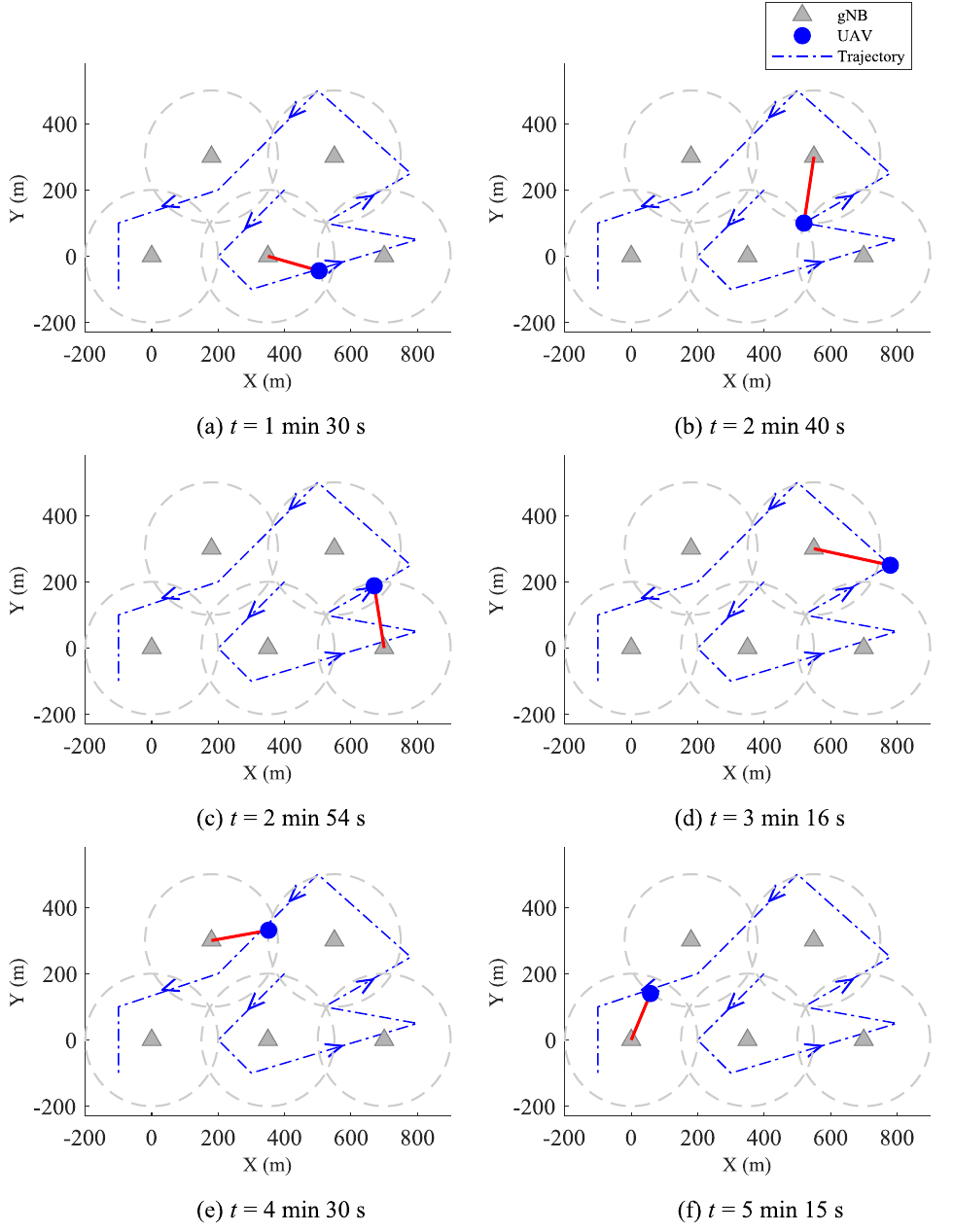}}
\caption{Visualization of  C-UAV system mobility and handover behavior in the simulation network. The six subfigures (a)–(f) represent snapshots along a fixed trajectory, where the UE dynamically switches base station connections. Notably, subfigures (b), (c), and (d) show the occurrence of ping-pong handovers when the UE moves near the boundary of two adjacent gNBs. }
\label{fig:handover}
\end{figure}

To verify the platform's comprehensive support for UAV mobility scenarios, this experiment strictly configures network node parameters (e.g., frequency bands) according to 3GPP standards and implements a full-process test including predefined trajectory planning, handover triggering and execution.
As shown in Fig. \ref{fig:handover}, five gNBs and one mobile UE emulating a UAV are deployed within a 1000 m × 600 m area. 
We also design a fixed flight path covering all base station service zones for the UAV and set the UE's cruising speed to 10–15 m/s (typical of consumer-grade drones), and a full trajectory duration of approximately 8 minutes.
The subfigure sequence captures the handover behavior during the UAV traversing coverage boundaries, where subfigures (b) to (d) clearly exhibit ping-pong effects at cell-edge regions. 
These results confirm the platform's capability to accurately simulate UAV mobility patterns, demonstrating its effectiveness in supporting dynamic UAV network processes. 

\subsection{Scalability}

\begin{figure}[htbp]
\centerline{\includegraphics[width=0.75\linewidth]{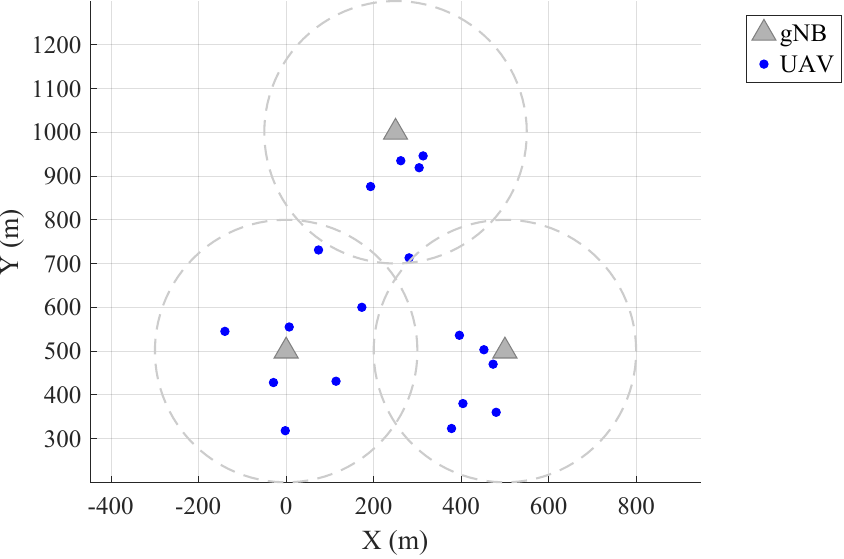}}
\caption{A network topology with three gNBs, each with seven UEs randomly distributed within their coverage areas.  }
\label{fig:5-gNBs}
\end{figure}

\begin{figure}[htbp]
\centerline{\includegraphics[width=1\linewidth]{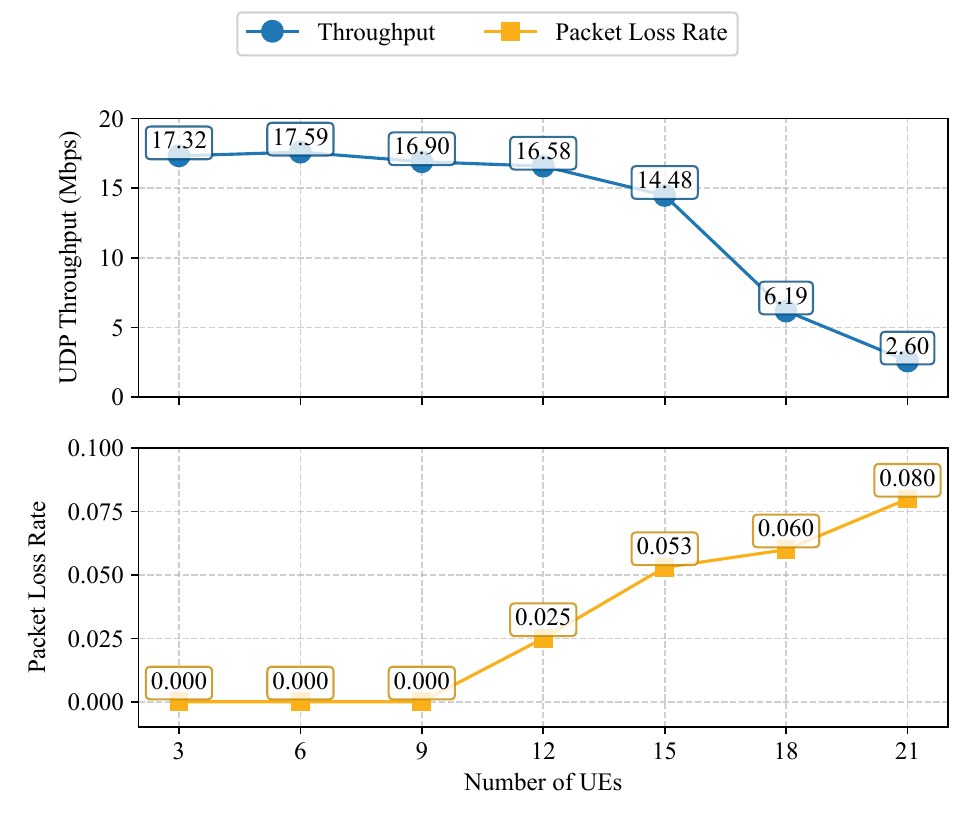}}
\caption{Network performance metrics with increasing UE density. }
\label{fig:multi-result}
\end{figure}

To validate the platform's scalability for massive device connectivity and demonstrate its architectural fitness for connected UAV swarm applications. 
As shown in Fig. \ref{fig:5-gNBs}, we configure an experiment featuring three gNBs and randomly distribute stationary UEs within their coverage areas.
UDP streams are used as external traffic sources to emulate UAV-generated data, such as video feeds or sensor telemetry. UDP provides a lightweight and continuous data flow, allowing us to realistically evaluate the platform’s performance under varying UE densities and network load.
Experimental results in Fig. \ref{fig:multi-result} show that individual UE throughput decreases as the number of UEs increases, while the packet loss rate remains below 0.1. 
This experiment confirms the platform's scalability. 

\subsection{Multiple Transport Layer Protocols Integration}

\begin{figure}[htbp]
\centerline{\includegraphics[width=1\linewidth]{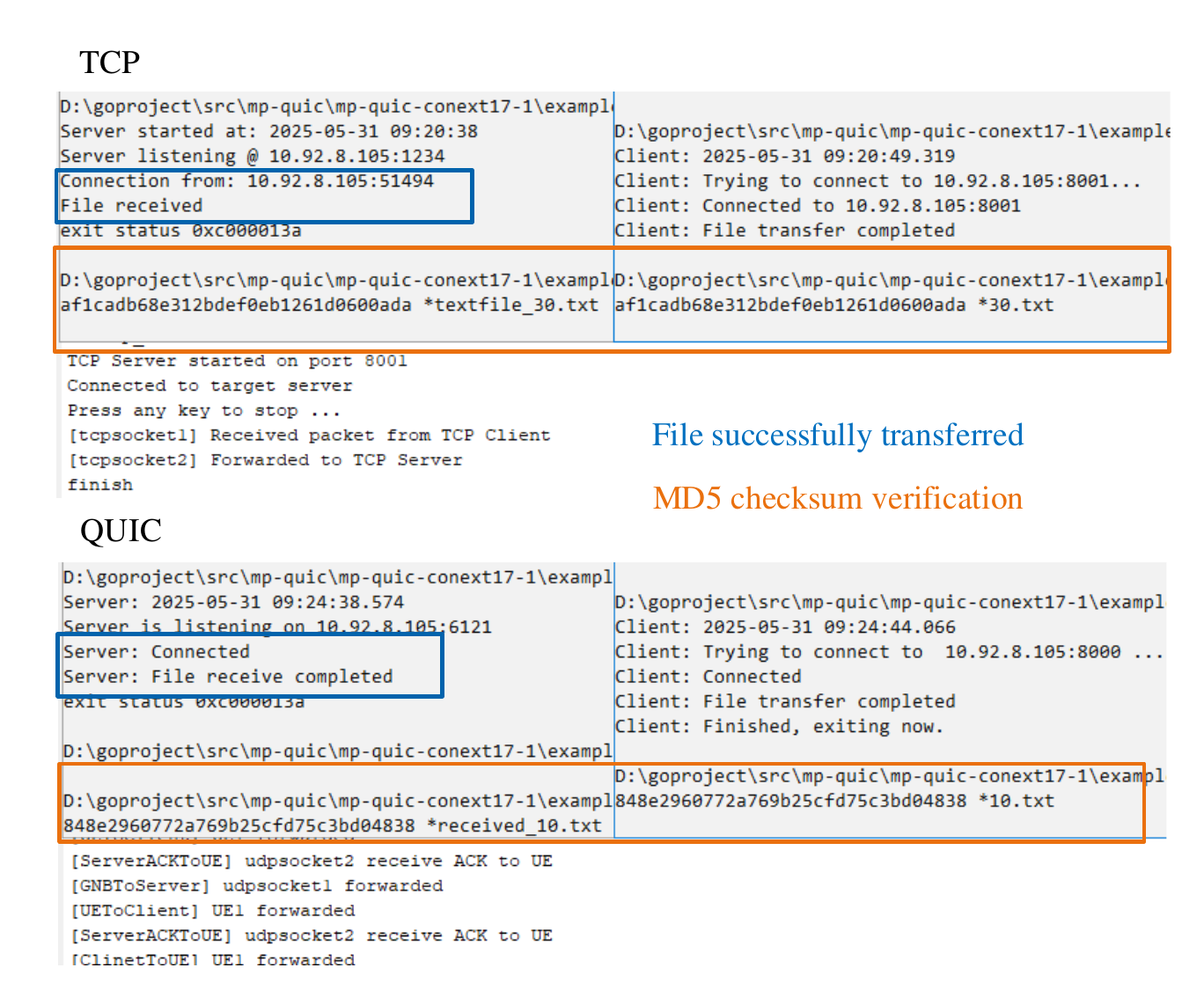}}
\caption{Protocol session trace for TCP and QUIC in simulation . }
\label{fig:result}
\end{figure}

To demonstrate the simulation platform’s support for multi-transport protocol integration, 
we conduct connectivity experiments using TCP and QUIC. 
In each experiment, a client-server pair connects via the MATLAB-simulated 5G network, with transport layer session establishment and end-to-end data exchange monitored through protocol-specific logging and trace visualization. 
After experiment, received files are validated using MD5 checksums to ensure data integrity. 
As illustrated in Fig. \ref{fig:result}, the upper gray sections depict terminal logs from the client and server, while the lower sections show MATLAB runtime logs. 
The results confirm that MATLAB effectively bridges packet transmission between the client and server, ensuring data traverses the simulated channels within our platform. This confirms successful independent activation, connection establishment, and real-time packet transmission for both protocols within the simulation environment, demonstrating the platform’s capability to integrate heterogeneous transport layer protocols. 

\subsection{Different Handover Strategies Comparison}

\begin{figure}[htbp]
\centerline{\includegraphics[width=1\linewidth]{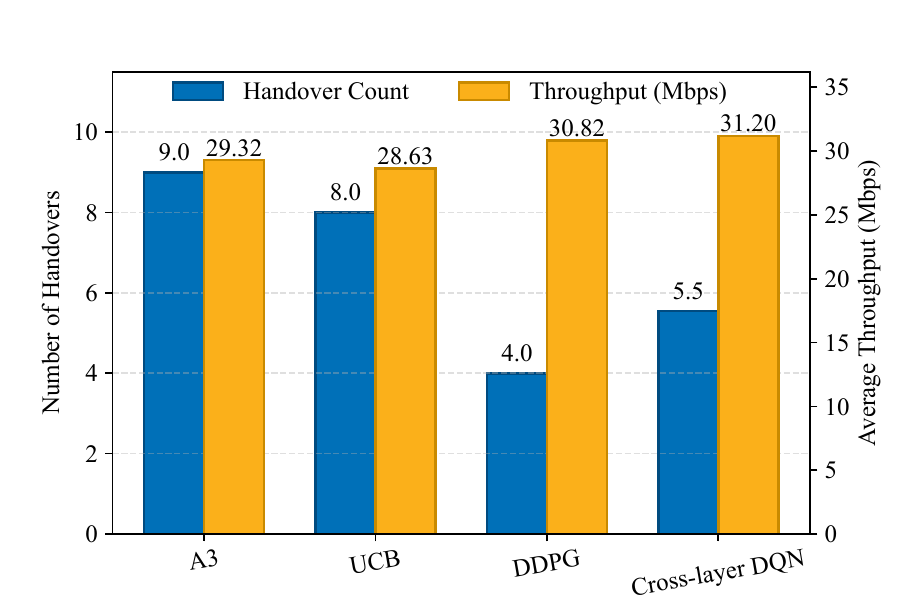}}
\caption{Average throughput and handover counts for different strategies. }
\label{fig:comparison}
\end{figure}

To validate the platform’s support for diverse handover management strategies and assess their effectiveness,
we compare four representative methods under the QUIC transport protocol: the traditional A3 event-based algorithm, a UCB-based online learning method\cite{sun2018smart}, a continuous control policy based on DDPG\cite{kwong2022ddpg}, and our proposed cross-layer DQN approach. 
The experiment adopts the identical 5-gNBs setup shown in Fig. \ref{fig:handover}. The DDPG and DQN algorithms undergo prior offline training within this environment using randomized trajectories distinct from evaluation paths. During testing, all four strategies are evaluated on the same set of 10 pre-generated random trajectories. Each algorithm executes 10 independent runs (one per trajectory, with each full trajectory requiring approximately 8 minutes to complete) from identical starting locations. Performance metrics, including handover count and end-to-end throughput are averaged across all runs. This methodology ensures fair comparison and assesses generalization capabilities. 
The experimental results are summarized in Fig. \ref{fig:comparison}.

Among the four handover strategies, the learning-based approaches demonstrate significantly fewer handovers compared to the traditional methods. This reduction can be attributed to the way learning algorithms evaluate long-term outcomes rather than immediate signal quality. Specifically, the DDPG method penalizes the ping-pong effect by integrating handover frequency into the reward function design. As a result, the agent learns to suppress unnecessary oscillatory handovers even if signal strength fluctuates temporarily, thereby improving handover robustness in highly dynamic environments.

In contrast, traditional methods are inherently greedy, triggering handovers as soon as a nearby base station exhibits a higher SINR. The A3-based strategy strictly follows the 3GPP-defined offset condition without foresight, while the UCB algorithm, as proposed in the SMART policy, selects the base station with the highest upper confidence bound of estimated throughput. Although UCB introduces exploration to avoid local optima, its reward design still emphasizes maximizing immediate throughput, leading to frequent handover events especially under fluctuating channel conditions.

In terms of throughput performance, the cross-layer DQN approach achieves the highest average throughput, followed closely by DDPG. Both methods incorporate throughput explicitly into their reward functions—DDPG uses SINR and instantaneous throughput as key indicators, while cross-layer DQN adopts a cross-layer design that captures PHY and Transport layer metrics in its state space. This allows the learning agent to select handover actions that not only minimize switching cost but also maximize sustained data rates over time.

Overall, the cross-layer DQN strategy demonstrates a favorable trade-off between reducing handover frequency and maintaining high throughput.
Importantly, this evaluation showcases the capability of our simulation platform to support learning-based handover strategies by providing access to multi-layer data, including physical-layer signal quality, transport-layer throughput, and historical mobility information.

\section{Conclusion}

This paper presents a modular simulation framework designed for investigating C-UAV communications within 5G networks. The developed platform enables scalable node configuration, supports the integration of diverse transport protocols, and facilitates comprehensive research into various handover strategies.

While the framework provides a robust foundation for simulating C-UAV systems, future work will focus on extending simulation capabilities and enhancing performance. Specifically, this includes incorporating more sophisticated air-to-ground channel models, exploring AI-driven optimization for handover and resource allocation, and accelerating simulations through parallel computing techniques. 

\section*{Acknowledgment}

This work was supported in part by the National Key R\&D Program of China under Grant 2024YFE0200700 and FDUROP (Fudan Undergraduate Research Opportunities Program)(Deang-Hui).
Thanks Dr. Yiling Yuan for insightful discussions on UAV channel model.

\vspace{12pt}
\bibliographystyle{IEEEtran}
\bibliography{references}

\end{document}